\documentclass[useAMS,usenatbib]{mn2e}
\pdfoutput=1
\usepackage{amsmath}
\usepackage{amssymb}
\usepackage{graphicx}
\usepackage{grffile}
\usepackage[usenames,dvipsnames]{color}

\usepackage{hyperref}

\newcommand{\Lya}{\ifmmode{{\rm Ly}\alpha}\else Ly$\alpha$\ \fi}

\newcommand{\ergps}{\,{\rm erg}\,{\rm s}\ifmmode{}^{-1}\else ${}^{-1}$\fi}

\newcommand{\dd}{\mathrm{d}}
\def\lsim{~\rlap{$<$}{\lower 1.0ex\hbox{$\sim$}}}
\def\gsim{~\rlap{$>$}{\lower 1.0ex\hbox{$\sim$}}}

\newcommand\aj{\rmfamily{AJ}}%
\newcommand\apj{\rmfamily{ApJ}}%
\newcommand\apjl{\rmfamily{ApJ}}%
\newcommand\aap{\rmfamily{A\&A}}%
\newcommand\mnras{\rmfamily{MNRAS}}%
\newcommand\nar{\rmfamily{New A Rev.}}%
\newcommand\pasa{\rmfamily{PASA}}%
\newcommand\nat{\rmfamily{Nature}}%
\title[Faint End Slopes of LAE and LBG LFs]{Connecting Faint End Slopes of the Lyman-$\alpha$ emitter and Lyman-break Galaxy Luminosity Functions}
\author[M. Gronke, M. Dijkstra, M. Trenti, S. Wyithe]{M. Gronke$^{1}$\thanks{E-mail:
maxbg@astro.uio.no}, M. Dijkstra$^{1}$, M. Trenti$^{2,3}$, S. Wyithe$^{3,4}$\\
$^{1}$Institute of Theoretical Astrophysics, University of Oslo, Postboks 1029, 0315 Oslo, Norway\\
$^{2}$Kavli Institute for Cosmology and Institute of Astronomy,
University of Cambridge, Cambridge, England\\
$^{3}$School of Physics, University of Melbourne, Parkville, VIC 3010, Australia\\
$^{4}$ARC Centre of Excellence for All-Sky Astrophysics (CAASTRO)
}

\begin{document}

\date{\today}

\pagerange{\pageref{firstpage}--\pageref{lastpage}} \pubyear{2015}

\maketitle

\label{firstpage}

\begin{abstract}
We predict Lyman-$\alpha$ (Ly$\alpha$) luminosity functions (LFs) of Ly$\alpha$-selected galaxies (Ly$\alpha$ emitters, or LAEs) at $z=3\--6$ using the phenomenological model of Dijkstra \& Wyithe (2012). This model combines observed UV-LFs of Lyman-break galaxies (LBGs, or drop out galaxies), with constraints on their distribution of Ly$\alpha$ line strengths as a function of UV-luminosity and redshift. Our analysis shows that while Ly$\alpha$ LFs of LAEs are generally not Schechter functions, these provide a good description over the luminosity range of $\log_{10}( L_{\alpha}/\ergps)=41-44$. Motivated by this result, we predict Schechter function parameters at $z=3-6$. Our analysis further shows that ({\it i}) the faint end slope of the Ly$\alpha$ LF is steeper than that of the UV-LF of Lyman-break galaxies, (with a median $\alpha_{Ly\alpha} < -2.0$ at $z\gsim 4$), and ({\it ii}) a turn-over in the Ly$\alpha$ LF of LAEs at Ly$\alpha$ luminosities $10^{40}$ erg s$^{-1}<L_{\alpha}\lsim 10^{41}$ erg s$^{-1}$ may signal a flattening of UV-LF of Lyman-break galaxies at $-12>M_{\rm UV}>-14$. We discuss the implications of these results -- which can be tested directly with upcoming surveys -- for the Epoch of Reionization.
\end{abstract}


\section{Introduction}
The luminosity function (LF) of galaxies provides one of the most basic statistical descriptions of a population of galaxies. It describes the number density of galaxies in a given luminosity interval. Generally, the LF is well described by a \citet*{Schechter1976ApJ...203..297S} function 
\begin{equation}
\phi(L)\dd L = \\\phi^* \left(\frac{L}{L^*}\right)^\alpha\exp\left(-\frac{L}{L^*}\right)\dd\left(\frac{L}{L^*}\right)
\label{eq:schechter}
\end{equation}
with a normalization parameter $\phi^*$, an exponential cutoff at $L\gtrsim L^*$, and, a power law with faint-end-slope $\alpha$ for $L\ll L^*$. The parameters depend on wavelength considered, galaxy type (e.g., passive versus star forming), and cosmic time. 

At high redshift, galaxies are typically identified either through their broadband colors, for example using the drop-out or Lyman-break technique \citep{Steidel1996ApJ...462L..17S}, or through narrow-band searches aimed at detecting emission lines \citep{Partridge1967ApJ...147..868P,Djorgovski1985ApJ...299L...1D}
In particular, young star forming galaxies emit a significant fraction of their radiation as Lyman-$\alpha$ (Ly$\alpha$) emission, and this method has been proved to be very efficient in finding samples out to $z\sim 7$ \citep[e.g.][]{2000ApJ...545L..85R,2001ApJ...563L...5R,Ouchi2008,2009ApJ...705..639B,2010ApJ...716L.200B,2010ApJ...714..255G,Kashikawa2011ApJ...734..119K,Hibon2012ApJ...744...89H,Ono2012ApJ...744...83O,Ota2012MNRAS.423..444O,2012ApJ...752L..28R,Shibuya2012ApJ...752..114S,2013Natur.502..524F,Konno2014arXiv1404.6066K}.

Galaxies that have been selected (found) on the basis of their Ly$\alpha$ lines are referred to as `Ly$\alpha$ emitters' (or LAEs). LAEs are useful because they are selected on having a strong Ly$\alpha$ line flux irrespective of their associated UV-continuum emission. Therefore, LAEs can be fainter in the continuum compared to Lyman-break galaxies, and complement galaxy samples obtained via broadband searches which have been extensively carried out with the Hubble Space Telescope out to $z\sim 10$ \citep[e.g.][]{2004ApJ...612L..93Y,2006AJ....132.1729B,2006ApJ...653...53B,2010MNRAS.403..938W,2011ApJ...727L..39T,2012A&A...547A..51G,2012ApJ...756..164F,Bouwens2014,2014arXiv1410.5439F,2014ApJ...786..108O,2014ApJ...786...57S}. Moreover, the sensitivity of the observed Ly$\alpha$ flux to intervening neutral hydrogen gas makes LAEs an excellent probe of the Epoch of Reionization \citep[see e.g.][for a review]{Dijkstra2014}.

Since the range of observed Ly$\alpha$ luminosities at high-$z$ typically extends only over $\sim 1\--1.5$ orders of magnitude, the shape of the Ly$\alpha$ LF is not strongly constrained and a fit with a Schechter function leads to significant degeneracy in the parameters. In particular the faint-end slope $\alpha_{\Lya}$ is essentially unconstrained: for example, \citet{Henry2012} used a sample of six (three) LAEs to find $\alpha_{\Lya}=-1.70^{+0.73}_{-0.57}$ ($\alpha_{\Lya}=-1.45^{+0.92}_{-0.70}$) at $z=5.7$. Other approaches include assuming a fixed value for $\alpha_{\Lya}$ and resorting to the data to constrain the other parameters \citep{2005MNRAS.359..895V,2007ApJ...671.1227D,Ouchi2008,Hu2010ApJ...725..394H,Kashikawa2011ApJ...734..119K, 2012ApJ...744..110C,2013MNRAS.431.3589Z}. In contrast, the ultraviolet (UV) LF of Lyman-break galaxies (LBGs) is much better constrained due to available data stretching over several orders of magnitude in luminosity \citep{UVLF2010MNRAS.403..960M,UVLF2011ApJ...728L..22Y,UVLF2011ApJ...737...90B,UVLF2012ApJ...760..108B,UVLF2012ApJ...759..135O,UVLF2012ApJ...761..177Y,UVLF2013MNRAS.429..150L,UVLF2013ApJ...768..196S}. For the faint end slope, the most recent by \citet{Bouwens2014} finds $\alpha_{UV}=-1.91\pm 0.09$ ($\alpha_{UV}=-1.64\pm 0.04$) at $z\sim 6$ ($z\sim 4$).

There exists a clear opportunity to connect LAEs and LBGs via the Ly$\alpha$ line emission properties of LBGs. \citet{Shapley2003ApJ...588...65S} provided a probability distribution function (PDF) of the rest-frame equivalent-width (EW) of the Ly$\alpha$ line in their sample of $\sim 800$ $z\sim 3$ LBGs. \citet{Dijkstra2012} showed that this observed PDF was well described by an exponential function, and that the characteristic scale-length of this function {\it increased} towards fainter UV-luminosities. While there do not exist equally well measured PDFs at higher redshifts and/or fainter UV-luminosities, recent studies have constrained both the redshift and UV-luminosity dependence of the so-called `Ly$\alpha$ fraction', which quantifies the fraction of LBGs for which the Ly$\alpha$ EW exceeds a certain value. The Ly$\alpha$ fractions  -- which represent integrated versions of the full EW PDF -- increase from $z=2$ to $z=6$ at fixed $M_{\rm UV}$ \citep{Stark2010MNRAS.408.1628S, Cassata2014arXiv1403.3693C} and from UV-bright to UV-faint galaxies \citep{Stark2010MNRAS.408.1628S,2011ApJ...728L...2S, 2011ApJ...743..132P,Ono2012ApJ...744...83O,2012ApJ...744..179S}. 

There have been several attempts to link the redshift evolution of LBGs and their Ly$\alpha$ fractions to LAE luminosity functions \citep{Dijkstra2012,Faisst2014ApJ...788...87F,Schenker2014}. In this paper, we follow the work of \citet{Dijkstra2012} and combine the most recent constraints on UV-LFs \& Ly$\alpha$ fractions to make predictions for \Lya LFs. \citet{Dijkstra2012} showed that this phenomenological model reproduces observed Ly$\alpha$ LFs and their redshift evolution remarkably well. Here, we focus specifically on the faint-end slope of the Ly$\alpha$ LF of LAEs, because ({\it i}) we can make robust predictions for this faint end slope, ({\it ii}) as we will show later, this faint end slope can be highly relevant for understanding the Epoch of Reionization.

This paper is structured as follows. In Sec.~\ref{sec:method}, we lay out our method. We present our results in Sec.~\ref{sec:results} and discuss them in Sec.~\ref{sec:discussion}. Finally, we conclude in Sec.~\ref{sec:conclusions}. The cosmological parameters we adopt are $\Omega_{\rm m}=0.3, \Omega_{\Lambda}=0.7, h=0.7, \sigma_8 = 0.9$.

\section{Method}
\label{sec:method}
The number density of LAEs with luminosities in the interval $[L_\alpha\pm\dd L_\alpha/2]$ is given by
\begin{equation}
\phi_{\rm LAE}(L_\alpha)\dd L_{\alpha} = \dd L_\alpha F \!\!\int\limits_{M_{UV,{\rm min}}}^{M_{UV, {\rm max}}}\!\! \dd M_{UV}\,\phi(M_{UV}) P(L_\alpha|M_{UV})
\label{eq:phiLalpha}
\end{equation}
Here, $\phi(M_{UV})\dd M_{UV}$ denotes the number density of LBGs as a function of in the range $M_{\rm UV}\pm \dd M_{\rm UV}/2$. This function can be represented by the Schechter function with parameters $(\alpha_{UV}, M_{UV}^*,\phi_{UV}^*)$. 

The term $P(L_\alpha|M_{UV})\dd L_\alpha$ is the conditional probability that a galaxy has a \Lya luminosity $L_{\alpha}$ given an absolute UV magnitude $M_{\rm UV}$. This conditional probability can be recast in terms of the equivalent width ($EW$) probability density function $P(EW|M_{UV})$ as $P(L_\alpha|M_{UV}) = P(EW|M_{UV})\frac{\partial EW}{\partial L_\alpha}$ if $EW > EW_{\rm LAE}$, where $L_{\alpha}$ and $EW$ are related as $L_{\alpha}=EW L_{\lambda}=EW [\nu L_{\nu}/\lambda]$. Here, the luminosity/flux densities, frequency and wavelength are evaluated just longward of the Ly$\alpha$ resonance at $\lambda=(1216+\epsilon)$\AA. We can extrapolate these flux/luminosities to their values where the UV-continuum measurements are usually made (see e.g. Dijkstra \& Westra 2010)\footnote{We use the relation $L_{\alpha}= C_1 EW L_{UV,\nu}$ where $L_{UV,\nu}\propto \nu^{-\beta-2}$ is the UV luminosity density $L_{UV,\nu}$ and $C_1 \equiv \nu_\alpha/\lambda_\alpha (\lambda_{UV}/\lambda_\alpha)^{-\beta-2}$ converts the flux density at $\lambda=(1216+\epsilon)$\AA\ to that at $\lambda_{UV}=1600$\AA, which is the wavelength where $L_{UV,\nu}$ was measured \citep{DW2010MNRAS.401.2343D}.}. Furthermore, $EW_{\rm LAE}$ denotes the equivalent width threshold that determines whether a galaxy would make it into an LAE sample. We adopt that $EW_{\rm LAE}=0$\AA, but note that some surveys adopt colour criteria for selecting LAEs as large as $EW_{\rm LAE}=64$\AA\ \citep[see][]{Dijkstra2012}. If $EW \le EW_{\rm LAE}$, then $P(L_\alpha|M_{UV}) = 0$ since in this case the galaxy does not qualify as an LAE.
This threshold more closely represents detection threshold for Ly$\alpha$ emitting galaxies in spectroscopic surveys\footnote{Note that in practise an EW cut is likely still needed to distinguish between LAEs and lower-$z$ interlopers, such as [OII] emitters. This EW cut can nevertheless be lower than $EW_{\rm LAE}\sim 20$\AA\ \citep{2015AAS...22533649L}} -- e.g., with \textit{MUSE} \citep{MUSE}, \textit{HETDEX} \citep{HETDEX} and/or \textit{VIMOS} \citep{Cassata2011A&A...525A.143C,Cassata2014arXiv1403.3693C}.
We have verified that our main results do not depend on this choice\footnote{We have verified that varying $EW_{\rm LAE}$ in the range $[0,\,50]\,$\AA\ changes $\alpha_{\Lya}$ by $\sim 0.02$.}.

The preceding factor $F$ in Eq.~\eqref{eq:phiLalpha} is merely a normalization constant to fit the data and, hence, can be thought of as the ratio of predicted versus the total number of LAEs. This factor should ideally be $F=1$. However, \citet{Dijkstra2012} required that $F\sim 0.5$. The origin of this number is not known \citep[see][for an extensive discussion]{Dijkstra2012}\footnote{The value of $F$ depends weakly on the adopted UV Schechter function parameters. For example, \citet{2014arXiv1411.2976B} reported slightly different best-fit values, which drive $F$ up to $F\sim0.7\--0.8$. The \citet{2014arXiv1410.5439F} parameters, on the other hand, also suggest $F\sim 0.5$.}, but we stress it only affects the predicted normalization linearly and not the predicted faint-end slopes.

Hence, the key function in our analysis is $P(EW|M_{UV})$. Several functional forms have been explored in the literature. \citet{Schenker2014} compared the maximum likelihood values for several EW distributions to their \textit{Keck MOSFIRE} \citep{KeckMosfire} data, and concluded that the exponential distribution introduced by \citet{Dijkstra2012} provides an adequate fit. This functional form is
\begin{equation}
P(EW|M_{UV},z) =\mathcal{N} \exp\left[-\frac{EW}{EW_{c}(M_{UV},z)}\right]
\label{eq:PEW_MUV}
\end{equation}
with 
$EW_{c}=EW_{c,0}+\mu_{M_{UV}} (M_{UV} + M_{UV,0}) + \mu_z (z + z_0)$
where $\mu_{M_{UV}}$, $\mu_z$, $M_{UV,0}$, $z_0$, and $EW_{c,0}$ are model parameters. These parameters were chosen to match the observations of \citep{Shapley2003ApJ...588...65S} and \citet{Stark2010MNRAS.408.1628S,2011ApJ...728L...2S} as closely as possible. Furthermore, $\mathcal{N}$ is a normalization constant which is forced to be zero outside of $[EW_{\rm min},\,EW_{\rm max}]$. Our choice of values for the model parameters is described in Sec.~\ref{sec:numerical-results} where we present the numerical results. In Appendix~\ref{sec:AppEWdist} we show explicitly that the main results in this paper are insensitive to both the functional form of $P(EW)$ and the parameterization of $EW_c$.

\section{Results}
\label{sec:results}
We first present results in which EW$_{\rm c}=$constant (in \S~\ref{sec:EWc_const}). This allows us to demonstrate that for models in which the Ly$\alpha$ fraction does not evolve with $M_{\rm UV}$, the faint end slope of the LF of LAEs approaches that of LBGs. We then present a simplified model in \S~\ref{sec:delta-func} in which the {\it mean} Ly$\alpha$ EW-PDF increases towards fainter UV-luminosity function. This model demonstrates quantitatively that the faint end slope of the LF of LAEs is steeper than that of LBGs if the Ly$\alpha$ fraction increases towards fainter UV-luminosities. In \S~\ref{sec:numerical-results} we present the results that we obtained from the EW-PDF given in Eq.~\eqref{eq:PEW_MUV}.

\begin{figure}
  \centering \includegraphics[width=\linewidth]{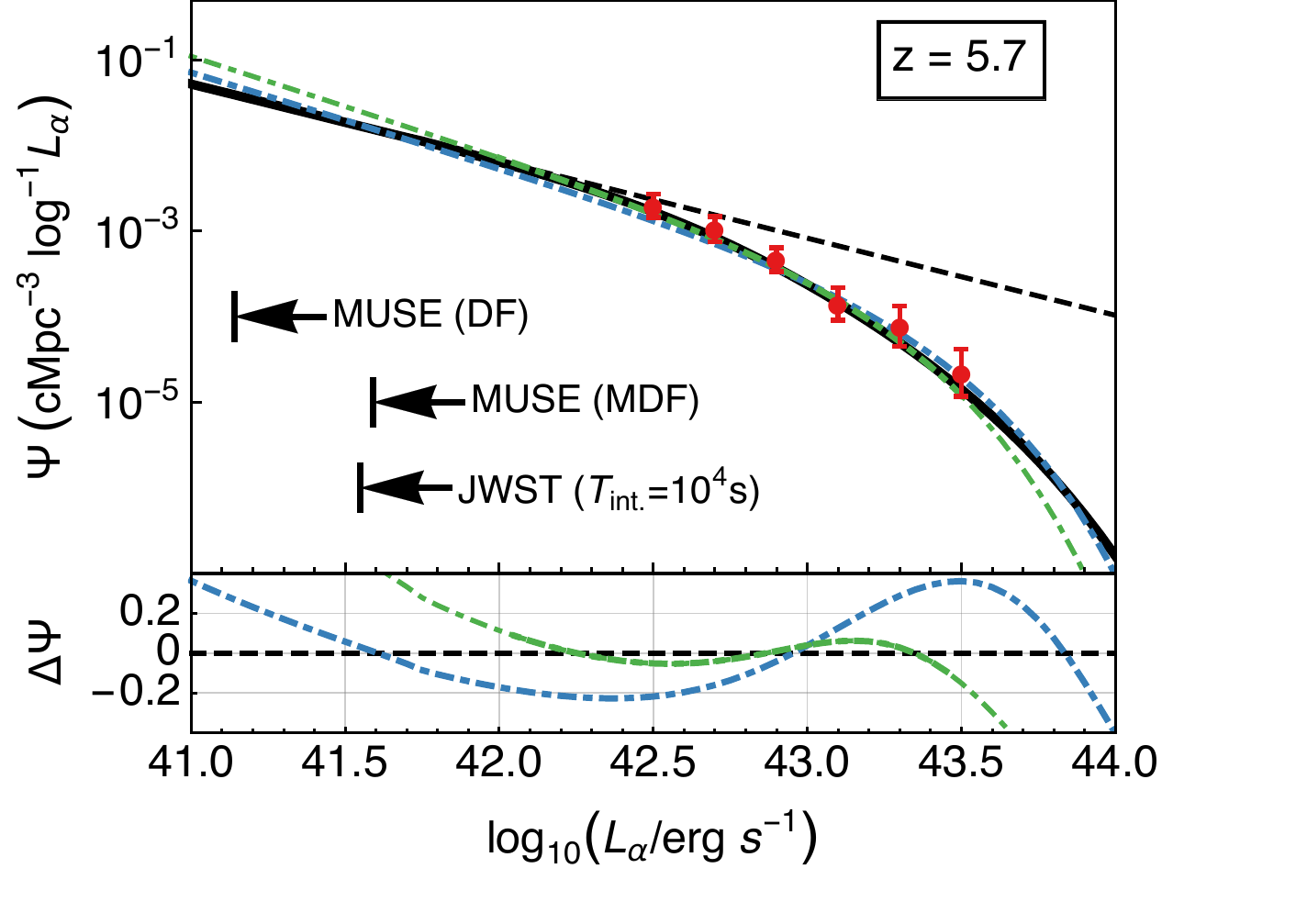}
  \caption{Upper panel: The predicted number of LAEs in the range $\log_{10}\left(L_\alpha/\ergps\right)\pm \dd\log_{10}\left(L_\alpha/\ergps\right)$ using the UV LF evolution from \citet{Bouwens2014} taken at $z=5.7$ (black solid line). The grey dashed line marks the faint end slope ($\alpha_{\Lya}=-1.90$) and the dashed dotted lines show Schechter fits to our numerical findings.
Once the  fit was carried out over the whole shown luminosity range (blue) and once only in $\log_{10}(L/\ergps) = [42, 43.5]$ (green). The red discs are the $z = 5.7$ observations by \citet{Ouchi2008} and the black arrows denote the \textit{MUSE} deep field and medium deep field as well as the \textit{JWST} limits at that redshift (see text for details). Lower panel: Relative deviation of the fits to the numerical results.}
\label{fig:LF}
\end{figure}

\subsection{Exemplary case with $EW_c = \mathrm{const.}$}
\label{sec:EWc_const}
We consider the case $P(EW|M_{UV}) = P(EW)$, i.e. $EW_c = \mathrm{const.} \equiv \lambda / C_1$. Furthermore, we set $\mathcal{N}=0$ for $EW<EW_{\rm LAE}$.
Under these assumptions we find
\begin{align}
  \phi(L_\alpha)\dd L_\alpha \propto &\; \int\limits_0^\infty L_{UV,\nu}^{\alpha-1}\exp\left[-\frac{L_{UV,\nu}}{L_{UV}^*} - \frac{L_\alpha}{\lambda L_{UV,\nu}}\right] \,\mathrm{d}L_{UV,\nu}\\
\propto &\; L_\alpha^{\alpha_{UV}/2} K_{-\alpha_{UV}}\left(2\sqrt{\frac{L_\alpha}{\lambda L_{UV}^*}}\right) \dd L_\alpha \label{eq:nalpha-expolast}
\end{align}
where $K_n(x)$ is the modified Bessel function of the second kind and $L_{UV}^*$ is the luminosity corresponding to $M^*_{UV}$.

Eq.~\eqref{eq:nalpha-expolast} shows that the \Lya LF generally does not take-on a Schechter form. The slope of the LF is given by
\begin{equation}
\alpha_{\Lya}\equiv\frac{\dd\log \phi(L_\alpha)}{\dd\log L_\alpha}=-\frac{\sqrt{y} K_{\alpha_{UV} - 1}(2\sqrt{y})}{K_{\alpha_{UV}}(2\sqrt{y})}
\end{equation}
with $y\equiv L_\alpha/(L_{UV}^* \lambda)$. For $L_\alpha \ll L_{UV}^* \lambda$ we have $y \ll 1$, and we obtain to leading order $\alpha_{\Lya}\approx -\Gamma(1-\alpha_{UV})/\Gamma(-\alpha_{UV}) = \alpha_{UV}$. Thus, having a {\it constant $EW_c$ corresponds to an unchanged faint end slope, $a_{\Lya} = \alpha_{UV}$}.

\begin{figure*}
  \centering
  \includegraphics[width=\linewidth]{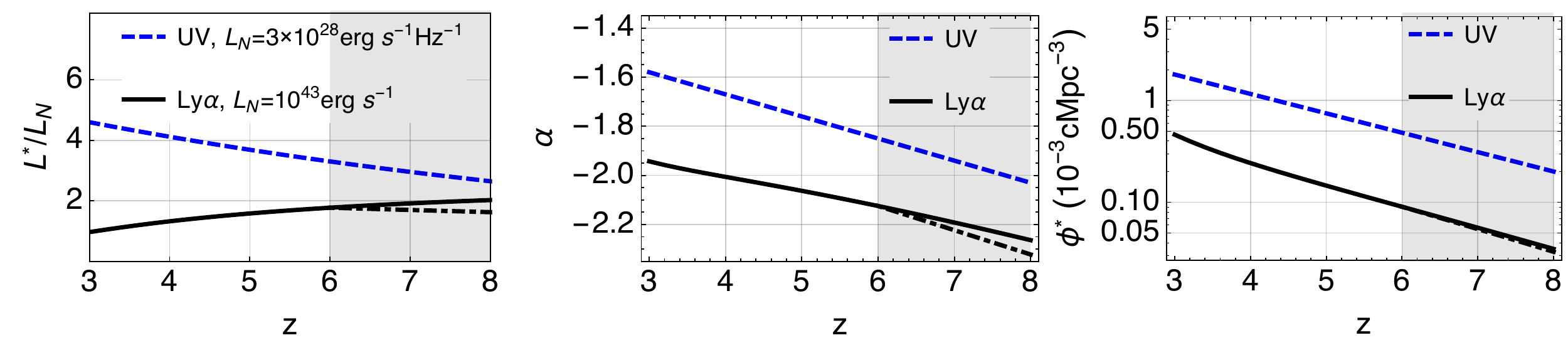}
  \caption{Evolution of the Schechter parameters of the UV LF from \citet{Bouwens2014} (as {\it blue dashed lines}) and the parameters of the computed Ly$\alpha$ LF ({\it black lines}). In particular, the {\it left panel} shows the characteristic luminosity $L^*$ (note the different normalization constants), the {\it central pane}l the faint end slope $\alpha$, and the {\it right panel} the overall normalization $\phi^*$. Predictions at $z>6$ (within the {\it shaded grey area}) do {\it not} account for reionization (see \S~\ref{sec:predgtrz6}). In this region, the {\it black solid lines} correspond to models with an uninterrupted EW evolution, whereas the {\it dashed-dotted lines} represent a model in which we freeze the EW evolution, i.e., $EW_c(z>6) = EW_c(z=6)$ (this assumption has been adopted in previous works).}
\label{fig:schechter_vs_z}
\end{figure*}

\subsection{Exemplary case where $P(EW|M_{UV})$ evolves with $M_{UV}$}
\label{sec:delta-func}
If $EW_c$ depends on $M_{UV}$ a general analytic solution for $\phi(L_{\alpha})$ does not exist. For illustration purposes we first consider a case in which we replace Eq.~\eqref{eq:PEW_MUV} with a Dirac-$\delta$ distribution, 
\begin{equation}
p(EW|L_{UV}) \propto \delta(EW - EW_d)\,,
\end{equation}
where $EW_d(L_{UV}) \equiv EW_{d,0} \left(L_{UV} / L_{UV}^*\right)^\gamma$. The parameter EW$_{\rm d}$ can be interpreted as the mean of the full PDF. This $\delta$-function PDF leads\footnote{For simplicity, we set the minimum and maximum UV luminosity to zero and infinity, respectively.} to 
\begin{multline}
\phi(L_\alpha)\dd L_\alpha \propto \dd L_\alpha\times L_\alpha^{\alpha_{UV}/(\gamma + 1)} \\
\times \exp\left[- \left( \frac{L_\alpha}{C_1 EW_{d,0} L_{UV}^*}\right)^{1/(\gamma + 1)} \right].
\end{multline}
Here, the faint end slope is $\alpha_{\Lya} = \alpha_{UV}/(\gamma + 1)$. 
he Ly$\alpha$ LF thus has a \textit{steeper} faint-end slope than the LBG LF, if $\gamma < 0$ (i.e. if $EW_d$ decreases towards fainter $L_{\rm UV}$, as has been observed).
Also note that we again obtain $\alpha_{\Lya} = \alpha_{UV}$ if $EW_d$ does not evolve with $M_{\rm UV}$.

\subsection{Realistic case with $P(EW|M_{UV})$ inferred from observations}
\label{sec:numerical-results}
For the model parameters of $P(EW|M_{\rm UV})$ in Eq.~\eqref{eq:PEW_MUV} we adopt the values from \citet{Dijkstra2012}\footnote{Specifically the model parameters related to $EW_{c}$ are given by $(EW_{c,0}, \mu_{M_{UV}}, \mu_z, M_{\mathrm{UV},0}, z_0, F) = (23\,{\rm \AA}, 7\,{\rm \AA}, 6\,{\rm \AA}, 21.9, -4.0, 0.53)$. 
The EW-PDF covers the range [$EW_{\rm min},EW_{\rm max}$]. Here, the lower limit $EW_{\rm min} \equiv -a_1$, where $a_1(M_{\mathrm{UV}})$ follows the form $a_1 = 20\,$\AA\ for $M_{UV}<-21.5$, $a_1=(20-6(M_{UV} + 21.5)^2)\,$\AA\ for $-21.5\le M_{UV}\le -19.0$ and $a_1=-17.5\,$\AA, otherwise (see Dijkstra \& Wyithe 2012). We used $EW_{\rm max}=1000\,$\AA\ but we verified that this choice does not affect our results quantitatively.\label{fn:model_params}}. Example EW-PDFs are shown in Appendix~\ref{sec:AppFiducialPEW}. For a more detailed motivation of this $P(EW)$ we refer the reader to \citet{Dijkstra2012}. 
We integrate the UV-LF over the range $M_{UV,{\rm (min,max)}} = (-30,\, -12)$ when predicting Ly$\alpha$ luminosity functions, and discuss the impact of varying $M_{\rm max}$ in Sec.~\ref{sec:discussion}.

The redshift evolution of the best fit Schechter parameters of the UV LF is taken from \citet{Bouwens2014} and given as $M_{UV}^* = -20.89 + 0.12 z$, $\phi_{UV}^* = 0.48\times 10^{-0.19 (z-6)} 10^{-3}{\rm cMpc}^{-3}$, and, $\alpha_{UV} = -1.85 - 0.09 (z-6)$. Following these analyses, we use $\lambda_{UV} = 1600\,$\AA\ as rest frame wavelength in which the UV continuum was measured and assume a UV spectral slope $\beta=-1.7$. This choice for $\beta$ does not affect our results (see Appendix~\ref{sec:more-realistic-beta} for detailed discussion).

The upper panel of Fig.~\ref{fig:LF} shows the resulting number density of LAEs at $z=5.7$ in the luminosity range $\log_{10} L_\alpha \pm \dd\log_{10}L_\alpha/2$, i.e., $\psi(L_{\alpha})\dd\log_{10} L_{\alpha}$, as a function of $L_\alpha$. This quantity is related to $\phi(L_{\alpha})$ as $\psi(L_\alpha) = \phi(L_\alpha)  L_\alpha \log 10$ (`$\log$' denotes the natural logarithm).
We compare these prediction to the data from \citet{Ouchi2008}. 
In addition, we show the \textit{MUSE} detection limits\footnote{\textit{MUSE} survey limits taken from \url{http://muse.univ-lyon1.fr/IMG/pdf/science_case_gal _formation.pdf}.} for its medium deep field (MDF, limiting flux $F>1.1\times 10^{-18}\ergps {\rm cm}^{-2}$, integration time $T_{\rm int.}=10$h), and, deep field (DF, $F>3.9\times 10^{-19}\ergps {\rm cm}^{-2}$, $T_{\rm int.}=80$h) surveys as well as an exemplary \textit{JWST}\footnote{\textit{JWST} survey limits obtained from \url{http://www.stsci.edu/jwst/science/sensitivity/}} limit ($F\gtrsim 10^{-18}\ergps {\rm cm}^{-2}$, $T_{\rm int.}=10^4$\,s).

Figure.~\ref{fig:LF} also shows two Schechter function approximations to our numerical findings fitted over the full luminosity-range shown (in {\it blue}) and over $\log_{10}(L_\alpha/\ergps) = [40.5, 42]$ (in {\it green}). Although we do not expect the resulting \Lya LF to be a Schechter function (as shown in Sec.~\ref{sec:delta-func}), it provides a reasonable fit over the displayed luminosity range. This can also be seen in the lower panel of Fig.~\ref{fig:LF}, where we display the relative deviation of the fits to the LF.

Fig.~\ref{fig:schechter_vs_z} shows the redshift evolution of the Schechter best fit parameters (as {\it black lines}). Predictions for $z>6$ do not account for reionization effects and are calculated with an unaltered EW evolution (solid line) as well as an EW-PDF which does not evolve after $z=6$ (dash-dotted line). We discuss this result separately in \S~\ref{sec:predgtrz6}. For comparison, we plot the corresponding redshift parameterization of the UV LF by \citet{Bouwens2014} (as \textit{blue dashed lines}). The {\it left panel} shows that $L^*$ increases by a factor $\sim 2$ over the redshift range $z=3\--6$, which differs from the redshift evolution in the characteristic UV-luminosity which drops by $~20\%$. This difference is driven by the redshift evolution in the Ly$\alpha$ EW-PDF, which in turn was inferred from the observed redshift-evolution of Ly$\alpha$ `fractions' over this redshift range. The {\it central panel} shows that $\alpha_\Lya < \alpha_{UV}$. This is again a consequence of inferred redshift evolution of the Ly$\alpha$-EW PDF (see \S~\ref{sec:delta-func}). This figure also illustrates the close-to-linear $\alpha_\Lya$--$z$ relation. This evolution is mostly driven by the redshift evolution of $\alpha_{UV}$. Finally, the {\it right panel} shows the predicted redshift evolution in $\phi^*$.

\begin{figure}
  \centering
  \includegraphics[width=\linewidth]{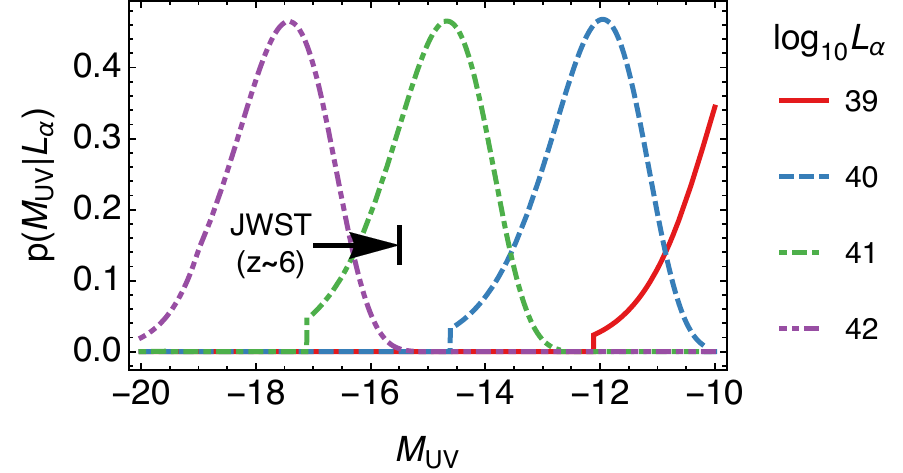}
  \caption{$p(M_{UV}|L_\alpha) \propto p(L_\alpha|M_{UV})\phi(M_{UV})$ versus $M_{UV}$ for some exemplary values of $L_\alpha$. The black arrow shows the \textit{JWST} photometric limit quoted by \citet{Windhorst2006NewAR..50..113W} for $z\sim 6$ which corresponds to $T_{\rm int.}\sim 10^6$\,s integration time.}
  \label{fig:pMUVLa}
\end{figure}

\section{Discussion}
\label{sec:discussion}

\subsection{Low-$L$ turnover}
\label{sec:low-l-turnover}
\begin{figure}
  \centering
  \includegraphics[width=\linewidth]{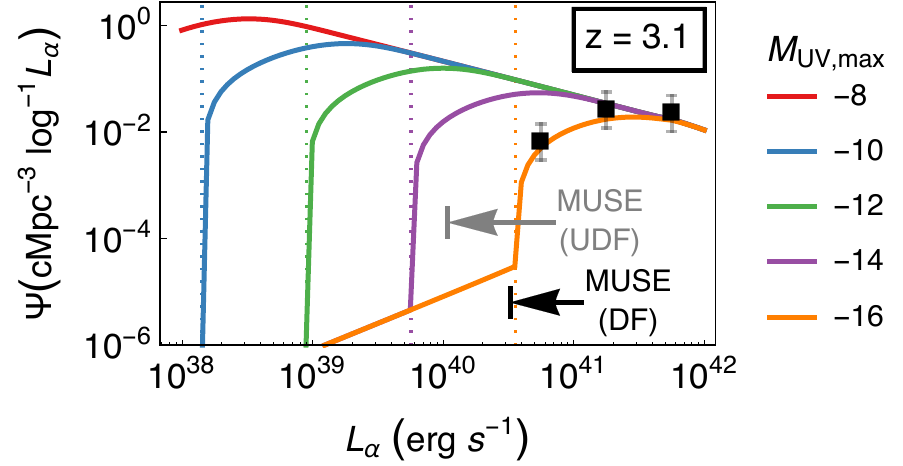}
  \caption{$\psi(L_\alpha)\dd L_\alpha$ versus $L_\alpha$ at $z=3.1$ for different values of $M_{UV,{\rm max}}$ illustrating the cutoff at low \Lya luminosities. 
The dashed and dotted lines show the ``cutoff'' and ``deviation'' points for each $M_{UV,{\rm max}}$ discussed in Sec.~\ref{sec:discussion}. Data points are from \citet{Rauch2008ApJ...681..856R} taken at that redshift (see the text for a discussion on the error bars). And the black (grey) arrow denotes planned future \textit{MUSE} (ultra) deep field limits.}
  \label{fig:low-l-bp}
\end{figure}

The integral over $\phi(L_\alpha)\dd L_\alpha$ diverges for $\alpha < -2$. We therefore expect that the luminosity function flattens or turns-over below some luminosity. The minimum luminosity that we can account for in our models is 
\begin{equation}
L_{\alpha, {\rm min}} = EW_{\rm min}(M_{\rm UV,max})\ C_1 L_{{\rm UV, min}},
\end{equation} where $EW_{\rm min}=-a_1=17.5$\,\AA\ (see footnote~\ref{fn:model_params} in \S~\ref{sec:numerical-results}) denotes the minimum equivalent width in our EW-PDF at the maximum absolute UV-magnitude (i.e. the lowest UV-luminosity). For example, we obtain $L_{\alpha, {\rm min}}\sim 10^{39}\ergps$ for $M_{\mathrm{UV}, {\rm max}}=-12$. At this luminosity we expect the predicted Ly$\alpha$ luminosity to go to zero, as is shown in Figure~\ref{fig:low-l-bp}. 

An estimate for where we may start to see departures from a power-law slope can be obtained by considering the conditional probability $p(M_{UV}|L_{\alpha})$. Bayes' theorem states that $p(M_{UV}|L_{\alpha}) \propto \phi(L_\alpha|M_{UV})\phi(M_{UV})$, of which we show examples in Fig.~\ref{fig:pMUVLa} for four different values of $L_{\alpha}$. This Figure illustrates for example that Ly$\alpha$ observations that probe a flux corresponding to $L_{\alpha}=10^{40}$ erg s$^{-1}$ -- a level that can be reached in \textit{MUSE} ultra deep fields -- effectively probe galaxies with $-14 < M_{\rm UV} < -11$, which are fainter than can be probed directly even with the \textit{JWST}. The \textit{JWST} detection limit shown in Figure~\ref{fig:pMUVLa} is taken from \citet{Windhorst2006NewAR..50..113W}. Figure~\ref{fig:pMUVLa} further shows that if the UV-LF flattens off at -- say -- $M_{\rm UV} \gsim -12$ that then the effects should become noticeable in the predicted Ly$\alpha$ luminosity function around $L_{\alpha}=10^{40}$ erg s$^{-1}$, as here galaxies with $M_{\rm UV} \sim -12$ dominate the contribution to the Ly$\alpha$ LF.

In Figure~\ref{fig:low-l-bp} we make these points more explicit, and show the predicted faint end of the LAE LF for four values of $M_{UV,{\rm max}}$ (calculated with the UV LF parameters at $z=3.1$). For each curve we marked $L_{\alpha, {\rm min}}$ with {\it dotted lines}. For example, a potential UV turnover at $M_{UV}\sim -12$ leads to deviations\footnote{The first deviations in the \Lya LF can be found at $L_{\alpha, {\rm dev.}} = C_1 L_{UV}(x) EW_c(x)$ with $x\equiv M_{UV, {\rm max}} - \Delta M_{UV}$. Here, $\Delta M_{UV}$ describes the half width of $p(M_{UV}|L_{\alpha})$ at a chosen probability threshold.} of the \Lya LF at $L_\alpha\sim 10^{40}\ergps$ and a cutoff at $L_\alpha \sim 10^{39}\ergps$. Figure~\ref{fig:low-l-bp} also contains data points taken from \citet{Rauch2008ApJ...681..856R}. \citet{Rauch2008ApJ...681..856R} performed an ultra deep ($92$-hr) exposure with \textit{VLT}s \textit{FORS2} low resolution spectrograph. The goal of these observations was to detect fluorescent Ly$\alpha$ emission from optically thick clouds powered by the ionizing background. While their sensitivity turned out not to be good enough to detect this fluorescent emission (revised estimates of the ionizing background and the conversion efficiency into Ly$\alpha$), they detected numerous ultra faint Ly$\alpha$ emitting sources characterizing their LF down to $L_\alpha\sim 6\times 10^{40}\ergps$. We computed the uncertainties with the cosmic variance calculator of \citet{Trenti2008ApJ...676..767T}. These data-points fall on the predicted LF for $M_{\mathrm{UV, max}}=-16$. 
However we caution that the turn-over occurs at the lowest luminosity data-point only, which might suffer from incompleteness (although it lies above the detection threshold).
In the same figure we provide the estimated \textit{MUSE} limits for the deep field (DF) and the gravitationally lensed ultra deep field (UDF) surveys.

\subsection{Implications for the Epoch of Reionization}
\label{sec:reion}
Low luminosity galaxies are expected to play a major role in driving the reionization of the Universe \citep[e.g.][]{Robertson2010,Trenti2010,  2012MNRAS.423..862K, 2014MNRAS.443L..44B}. Determining the faint end of the LF such as its slope and a turnover luminosity is essential for constraining the volume emissivity of ionizing photons. However, even future experiments will have difficulties detecting these galaxies directly via their UV continuum flux. Current constraints rely, therefore, on extrapolation of local properties to higher redshifts \citep{Weisz2014}, (relatively few) gravitationally lensed objects \citep{Alavi2014ApJ...780..143A,2014ApJ...786...60A} or inferences from gamma-ray burst observations \citep{Trenti2012ApJ...749L..38T}. In this work, we have shown that the \Lya LF can provide an independent probe of the faint end of the UV LF, and that for example the \textit{MUSE} DF survey could already detect (or rule out) a turnover at $M_{UV}\lesssim -15$.

Recent studies have shown that \Lya escape may be correlated with the escape of ionizing photons \citep{Behrens2014,Verhamme2014}, as the escape of ionizing photons requires low HI-column density ($N_{\rm HI} < 10^{17}$ cm$^{-2}$) channels, which can also provide escape routes for Ly$\alpha$ photons. The fact that Ly$\alpha$ LFs are likely steeper than the UV LFs implies that the Ly$\alpha$ volume emissivity -- and therefore possibly the ionizing emissivity -- are weighted more strongly towards low luminosity galaxies. This is consistent with the expectation that ionizing photons escape more easily from lower mass -- and hence lower luminosity -- galaxies.  A steep faint-end slope of the Ly$\alpha$ LF may therefore provide observational support for this scenario.

\subsection{Predictions for redshifts $z=6\-- 8$}
\label{sec:predgtrz6}
We extrapolated our predictions for the best-fit Schechter parameters of the LAE LF to $z>6$ in two ways (shown in Fig.~\ref{fig:schechter_vs_z}): \textit{(i)} in the first, we assume that the EW-PDF continues to evolve as inferred from the observations at $z=3\--6$. This model is represented by the {\it solid lines}, and, \textit{(ii)} in the second, we `freeze' the EW distribution for $z>6$ at the value it had at $z=6$ ({\it dashed lines}). This latter assumption has been common in previous works \citep[see e.g.][]{Dijkstra2011MNRAS.414.2139D,2013MNRAS.429.1695B,Jensen2013MNRAS.428.1366J,2014arXiv1412.4790C,Mesinger2014arXiv1406.6373M}. We show results for these two models to get a sense for the uncertainties on our predictions. We stress that we have purposefully {\it not} modelled the impact of reionization on the EW-PDF. Reionization is likely responsible for the observed `drop' in the observed Ly$\alpha$ fractions at $z>6$ 
\citep[e.g.][]{2011ApJ...743..132P,2012ApJ...744..179S,Ono2012ApJ...744...83O,Treu2013ApJ...775L..29T,Caruana2014MNRAS.443.2831C,Tilvi2014ApJ...794....5T}. Understanding this drop has been the main focus of previous works, and is outside the scope of this paper. Our predictions for \mbox{$z=6\-- 8$} are useful in a different way, as they provide predictions for the Ly$\alpha$ LFs of LAEs in the absence of reionization. Comparison to observed LFs at these redshifts highlight the impact of reionization.

\section{Conclusions}
\label{sec:conclusions}
We predicted Ly$\alpha$ luminosity functions (LFs) of Ly$\alpha$-selected galaxies (Ly$\alpha$ emitters, or LAEs) at $z=3\--6$ using the phenomenological model of \cite{Dijkstra2012}. This model combines observed UV-LFs of Lyman-break galaxies (LBGs), with observational constraints on the Ly$\alpha$ EW PDF of these LBGs, as a function of $M_{\rm UV}$ and redshift. The results from our analysis can be summarized as follows:
\begin{itemize}

\item While Ly$\alpha$ luminosity functions of LAEs are generally not Schechter functions, these provide a good description over the luminosity range of $\log_{10}( L_{\alpha}/\ergps)=41-44$ (see Fig.~\ref{fig:LF}). 

\item We predict Schechter function parameters at $z=3-6$ (shown in Fig.~\ref{fig:schechter_vs_z}). The faint end slope of the Ly$\alpha$ LF is steeper than that of the UV-LF of LBGs, with a median $\alpha_{Ly\alpha} < -2.0$ at $z\gsim 4$ (see the {\it central panel} in Fig.~\ref{fig:schechter_vs_z}). While the current work was in the advanced stage of completion, \citet{Dressler2014arXiv1412.0655D} posted a preprint in which they observationally infer a very steep faint end slope at $z\sim 5.7$ ($-2.35 < \alpha < -1.95$, also see \citet{Dressler2011ApJ...740...71D}). The central value $\alpha=-2.15$ is in excellent agreement with the value $\alpha\sim -2.1$ predicted in our framework.

\item The faint end of the LAE LF provides independent constraints on the very faint end of the UV-LF of LBGs. For example, the predicted LAE LF at Ly$\alpha$ luminosities $10^{40}$ erg s$^{-1}<L_{\alpha}\lsim 10^{41}$ erg s$^{-1}$ is sensitive to the UV-LF of LBGs in the range $-11>M_{\rm UV}>-15$ (see Fig.~\ref{fig:pMUVLa} and Fig.~\ref{fig:low-l-bp}). These LBGs are too faint to be detected directly (even with JWST). A turn-over in the Ly$\alpha$ LF of LAEs may signal a flattening of UV-LF of LBGs. We discuss implications of these results for the Epoch of Reionization in \S~\ref{sec:reion}.
 \end{itemize}

We have verified that these results are insensitive to our assumed functional form of $P(EW)$ and how we parameterized its dependence on $z$ and $M_{\rm UV}$. Our predictions can be tested directly with various upcoming surveys.

\section*{Acknowledgements}
We thank Masami Ouchi for kindly providing the data points shown in Fig.~\ref{fig:LF}. MD and MT thank Alan Dressler giving a presentation which inspired this work at the UCSB GLASS meeting in May 2014.
\bibliography{references}

\appendix
\section{Varying the EW distribution}
\label{sec:AppEWdist}

In this appendix, we demonstrated that our main results and conclusions do not depend on our assumed EW-PDF.

\subsection{Fiducial $P(EW)$}
\label{sec:AppFiducialPEW}
\begin{figure}
  \centering
  \includegraphics[width=\linewidth]{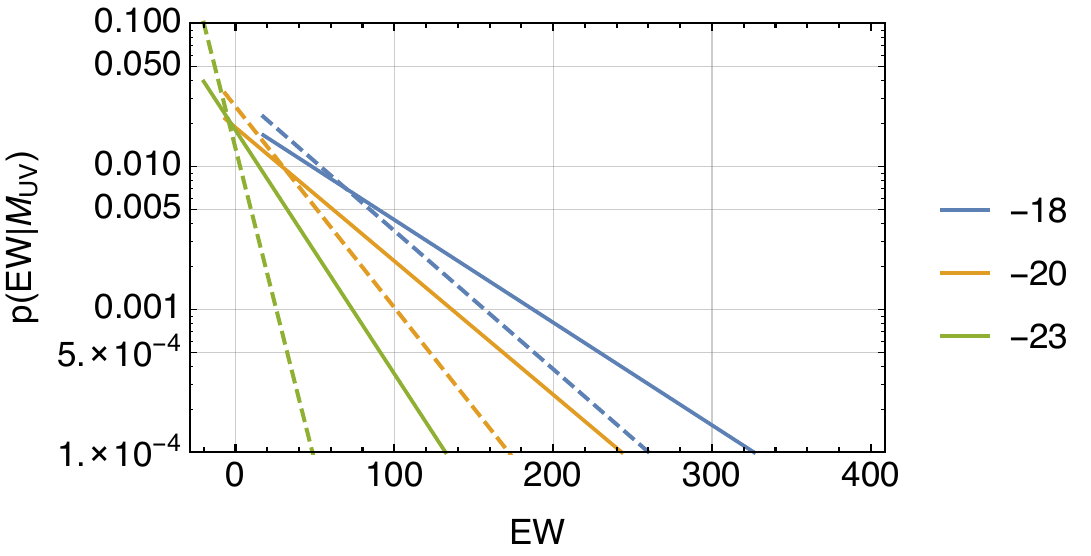}
  \caption{The PDF $P(EW|M_{{\rm UV},z})$ as described in Eq.~\eqref{eq:PEW_MUV} for three different values of $M_{\rm UV}$. The solid (dashed) lines shows the EW distribution for $z=5.7$ ($z=3.1$). Negative EWs corresponds to having Ly$\alpha$ in absorption. Note that for about half of $z\sim 3$ LBGs, Ly$\alpha$ is seen in absorption \citep{Shapley2003ApJ...588...65S}. Our formalism incorporates this measurement \citep[see][]{Dijkstra2012}.}
  \label{fig:pEW_MUV}
\end{figure}

Our default EW distribution is given by Eq.~\eqref{eq:PEW_MUV}. We plot the PDF for three UV magnitudes and two redshifts in Fig.~\ref{fig:pEW_MUV}.

\subsection{\citet{Schenker2014} parameterization}
\label{sec:schenk-param}
As mentioned in Sec.~\ref{sec:method}, \citet{Schenker2014} suggest an alternative parameterization of the EW-PDF, namely
\begin{equation}
P(EW|\beta) = \frac{A_{\rm em}}{\sqrt{2 \pi} \sigma EW} \exp\left[-\frac{(\log EW - \mu(\beta))^2}{2 \sigma^2}\right]\;.
\label{eq:PEW_schenker}
\end{equation}
This log-normal PDF possesses the parameters $A_{\rm em}$, $\sigma$ and $\mu$. The latter is given by
\begin{equation}
\mu(\beta) = \mu_\alpha + \mu_s (\beta - 2.0),
\end{equation}
where $\beta$ is the UV continuum slope. \citet{Schenker2014} found their EW distribution to depend more strongly on $\beta$ than on $(M_{\rm UV}, z)$, and therefore constrained $P(EW|\beta)$. 
We can include this parameterization into our formalism if we map $P(EW|\beta)$ onto $P(EW|M_{\rm UV},z)$.

This mapping is based on three results from \citet{2014ApJ...793..115B}:
\begin{enumerate}
\item We use their empirical linear correlation between $\beta\--M_{\rm UV}$ at $z\sim 7$. This relation constrains $\beta(M_{\rm UV} = -19.5) = -2.05\pm 0.09\pm 0.13$.
\item Furthermore, we apply their measured change per unit redshift $\Delta\beta/\Delta z = -0.1\pm 0.05$.
\item Finally, we use their measurement that $\Delta\beta / \Delta M_{\rm UV} = -0.2$ ($-0.08$) for $M_{\rm UV}\le -19$ ($>-19$).
\end{enumerate}
Accordingly, our mapping can be written as
\begin{equation}
\beta=\beta_0+\mu^{(\beta)}_{M_{\rm UV}}(M_{\rm UV}+19)+\mu^{(\beta)}_z(z-7),
\end{equation}
where $\mu^{(\beta)}_{M_{\rm UV}}\equiv \Delta \beta/\Delta M_{\rm UV}$ and $\mu^{(\beta)}_z\equiv \Delta \beta/\Delta z$.

\begin{figure}
  \centering
  \includegraphics[width=\linewidth]{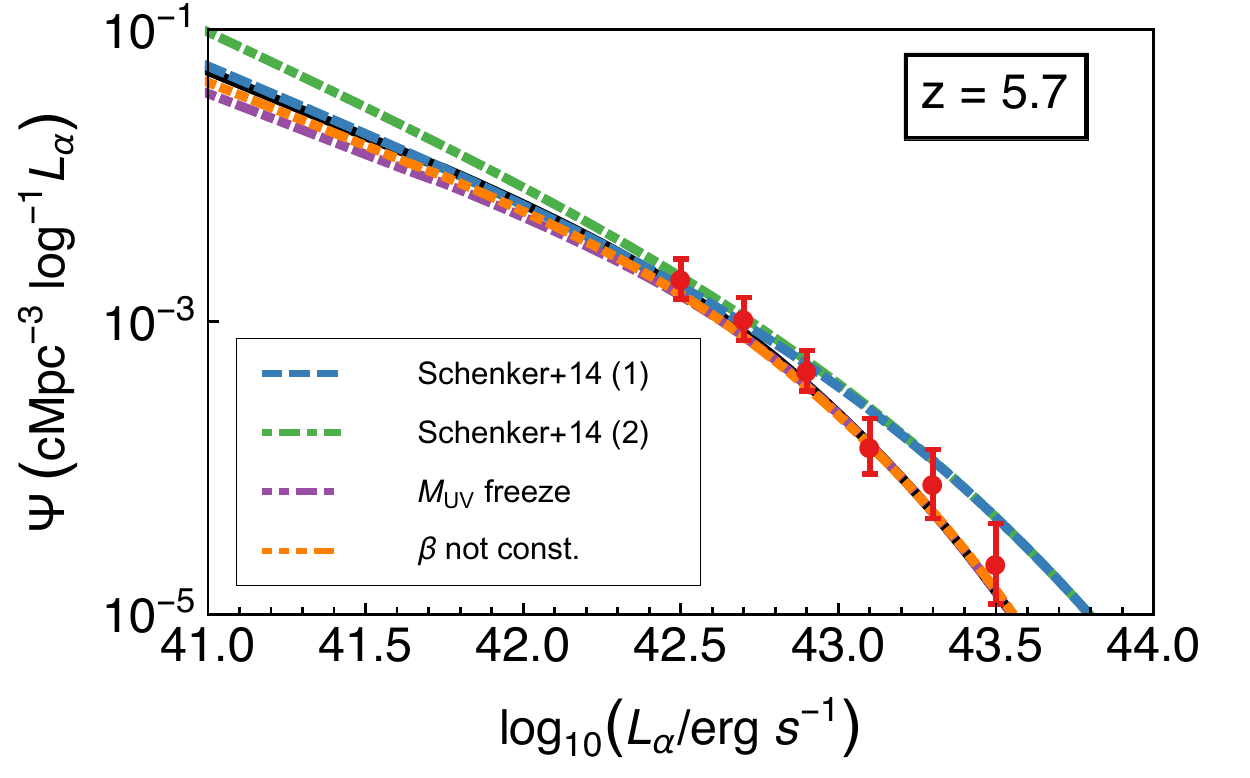}
  \caption{The \Lya LF at $z=5.7$ for different EW distributions. The solid black line and \citet{Ouchi2008} data points are the same as shown in Fig.~\ref{fig:LF}. As comparison we show the \Lya LF computed using the \citet{Schenker2014} parameterization of the EW PDF once with the full $\beta\--M_{\rm UV}$ relation as given in \S~\ref{sec:schenk-param} (blue dashed line) and once with a constant $\mu^{(\beta)}_{M_{\rm UV}} = -0.2$ over all $M_{\rm UV}$ (green dashed-dotted line). Also shown are our results when freezing the EW PDF at $M_{\rm UV}=-19$ (purple) and when using the $\beta\--M_{\rm UV}$ relation instead of a constant $\beta$ (orange).}
  \label{fig:LyaLF-diffEW}
\end{figure}

In Eq.~\eqref{eq:PEW_schenker} we used the best parameters by \citet{Schenker2014}, i.e., $(A_{\rm em},\,\sigma,\,\mu_\alpha,\,\mu_s)=(1.0,\,1.3\,,2.875,\,-1.125)$. In addition, since $A_{\rm em}$ is degenerate with $F$, we set $F=1$. The orange dashed line in Fig.~\ref{fig:LyaLF-diffEW} shows the resulting \Lya LF at $z=5.7$. The agreement in the faint-end between the two procedures is remarkable. For greater luminosities, however, the \citet{Schenker2014} parameterization leads to a (much) higher number density of LAEs. While there are significant uncertainties in the above procedure, the agreement we get at the faint end slope is especially encouraging. Future surveys can be extremely useful in further connecting the LAE and LBG populations by constraining the bright end of the LAE luminosity function.

\subsection{`Freezing' the EW PDF for faint galaxies}
\label{sec:freezing-pEW}
Since the evolution in the EW PDF for fainter sources involves a (modest) extrapolation of observationally inferred $P(EW)$, we have also tested and alternative PDF where we `froze' the evolution at $M_{\rm UV}=-19$. That is, we also conservatively assume that the EW-PDF stops evolving at $M_{UV} > -19$ (even though observations hint that this is not the case, see fig.~13 of \citet{Stark2010MNRAS.408.1628S}). Fig.~\ref{fig:LyaLF-diffEW} shows the resulting \Lya LF (purple line). It is clear that our results are only affected slightly, i.e., the faint-end-slope $\alpha_{\Lya}$ is reduced by $\sim 0.05$. We also tested this over a variety of redshifts.

\subsection{Non-constant UV spectral slope}
\label{sec:more-realistic-beta}
The spectral slope $\beta$ is not a constant, but depends on UV magnitude and redshift (as discussed above). This introduces  some additional dispersion in the predicted Ly$\alpha$ flux at a fixed $M_{\rm UV}$. However, varying $\beta$ within $[-2.0,\,-1.5]$ changes the \Lya flux only by $1 - (\lambda_{\rm UV} / \lambda_\Lya)^{1.5 - 2.0}\sim 13\%$. This dispersion is smaller than that introduced by the EW-PDF. If we replace the constant $\beta$ with the empirical fit described in \S~\ref{sec:schenk-param},  then our predicted \Lya LF (represented by the {\it orange line} in Fig.~\ref{fig:LyaLF-diffEW}) is barely any different from our fiducial model that used $\beta = -1.7$ (represented by the {\it black solid line}). 

\label{lastpage}
\end{document}